\documentclass[aps,twocolumn,showpacs,amsmath,amssymb,citeautoscript]{revtex4-1}

\usepackage{graphicx}% Include figure files
\usepackage{dcolumn}% Align table columns on decimal point
\usepackage{bm}% bold math
\usepackage{color}
\usepackage{enumerate}

\newlength{\figwidth}
\begin{document}

\title{Quantum critical point lying beneath the superconducting dome in iron-pnictides}

\author{T.~Shibauchi$^1$}
\author{A.~Carrington$^2$}
\author{Y.~Matsuda$^1$}

\affiliation{
$^1$Department of Physics, Kyoto University, Kyoto 606-8502, Japan\\
$^2$H.\,H. Wills Physics Laboratory, University of Bristol, Tyndall Avenue, Bristol, BS8 1TL, U.K.
}

\date{\today}

\begin{abstract}
{\bf
Whether a quantum critical point (QCP) lies beneath the superconducting dome has been a long-standing issue that remains unresolved in many classes of unconventional superconductors, notably cuprates, heavy fermion compounds and most recently iron-pnictides.  The existence of a QCP may offer a route to understand: the origin of their anomalous non-Fermi liquid properties, the microscopic coexistence between unconventional superconductivity and magnetic or some exotic order, and ultimately the mechanism of superconductivity itself.  The isovalent substituted iron-pnictide BaFe$_2$(As$_{1-x}$P$_x$)$_2$ offers a new platform for the study of quantum criticality, providing a unique opportunity to study the evolution of the electronic properties in a wide range of the phase diagram.  Recent experiments in BaFe$_2$(As$_{1-x}$P$_x$)$_2$ have provided the first clear and unambiguous evidence of a second order quantum phase transition lying beneath the superconducting dome.
}
\vspace{5mm}


\end{abstract}

\maketitle
\section{Introduction}
The discovery of iron-pnictide high-$T_c$ superconductivity has been one of the most exciting recent developments in condensed matter physics.  In 2006, Hideo Hosono's research group found superconductivity below 6\,K in LaFePO \cite{Kamihara2006}.  They discovered that by replacing phosphorus with arsenic and doping the structure by substituting some of the oxygen atoms with fluorine they could increase $T_c$ up to 26\,K \cite{Kamihara2008}.  This high $T_c$ in LaFeAs(O,F)  aroused great interest  in the superconducting community, particularly when it was found that $T_c$ could be increased up to 43\,K with pressure \cite{Takahashi2008}.  By the end of April 2008, it was found that $T_c$ could be increased to 56\,K by replacing La with other rare earth elements \cite{Wang2008}. Thus iron-pnictides joined the cuprates and became a new class of \emph{high-$T_c$} superconductor.

The most important aspect of the iron-pnictides may be that they open a new landscape in which to study mechanisms of unconventional  pairing  which lead to high-$T_c$  superconductivity \cite{Ishida2009,Paglione2010,Stewart2011,Mazin2010,Hirschfeld2011}.  The high transition temperatures in both cuprates and iron-pnictides cannot be explained theoretically by the conventional electron-phonon pairing mechanism and thus there is almost complete consensus that the origin of superconductivity of both systems has an unconventional origin \cite{Scalapino1995,Hirschfeld2011}.  Another class of materials in which there is extensive evidence for unconventional superconductivity are the heavy fermion compounds \cite{Pfleiderer2009}.  The unusual properties of these materials originate from the $f$ electrons in the Ce ($4f$) or U ($5f$) atoms which interact with the conduction electrons to give rise to heavy effective electron masses (up to a few hundred to a thousand times the free electron mass) through the Kondo effect.

There are several notable similarities between these three classes of unconventional superconductor. First of all, it is widely believed that in all three systems electron correlation effects play an important role for the normal-state electronic properties as well as the superconductivity. As in high-$T_c$ cuprates and some of the heavy fermion compounds, superconductivity in iron-pnictides emerges in close proximity to an antiferromagnetic (AFM) order, and $T_c$ has dome-shaped dependence on doping or pressure. In these three systems near the optimal $T_c$ composition various normal-state quantities often show a striking deviation from conventional Fermi liquid behavior.

Structurally, iron-pnictides also have some resemblance to cuprates: pnictides are two dimensional (2D) layered compounds with alternating Fe-pnictogen (Pn) layers sandwiched between other layers which either donate charge to the Fe-Pn layers or create internal pressure. However, there are also significant differences between three systems.  For example, the parent compounds of the iron-pnictides are metals whereas for cuprates they are Mott insulators.  Moreover, whereas in cuprates the physics is captured by single band originating from a single $d$-orbital per Cu site, iron-based superconductors have six electrons occupying the nearly degenerate $3d$ Fe orbitals, indicating that the system is intrinsically multi-orbital and therefore that the inter-orbital Coulomb interaction also plays an essential role.  Indeed, it is thought that orbital degrees of freedom in pnictides give rise to a rich variety of phenomena, such as nematicity and orbital ordering \cite{Yildirim2009,Ku2009,Chen2010,Chuang2010,Chu2010,Shimojima2010,Yi2011,Kim2011,Fernandes2011,Fernandes2012,Fernandes2012a,Kontani2011a,Onari2012b,Bascones2012,Phillips2012,Kasahara2012,Blomberg2013}.
In cuprates a crucial feature of the phase diagram is the mysterious pseudogap phase \cite{Timusk1999,He2011,Vishik2012,Lawler2010,Daou2010,Fauque2006,Xia2008,Wu2011}.  At present it is unclear if an analogous phase exists in iron-pnictides.

In heavy fermion compounds, the $f$ electrons, which localize at high temperature, become itinerant at low temperature through Kondo hybridization with the conduction electrons.  Heavy fermion compounds usually have complicated 3D Fermi surfaces.  The competition of various interactions arising from Kondo physics often makes their magnetic structures complicated.  Orbital physics is also important in heavy fermion compounds, as shown by multipolar ordering (this corresponds to orbital ordering in $d$ electron system) \cite{Kuramoto2009}, but often its nature is not simple due to the complicated Fermi surface and strong spin-orbit interaction.   Iron-pnictides, in sharp contrast,  have much simpler quasi-2D Fermi surface with weaker spin-orbit interaction and simple magnetic structures \cite{Dai2012}.

A quantum critical point (QCP) is a special class of second order phase transition that takes place at absolute zero temperature, typically in a material where the phase transition temperature has been driven to zero by non-thermal parameters such as doping and the application of pressure or magnetic field \cite{Sachdev,Vojta2003,Sachdev2011,Sondhi1997,Varma2002,Lohneysen2007}. In this short review we will address several questions concerning the physics of a putative QCP in the phase diagram of the iron-pnictide materials \cite{Abrahams2011}. In particular, we will review the evidence for a QCP hidden beneath the superconducting dome, which we believe to be crucially important for understanding the anomalous normal-state properties and the high-$T_c$ superconductivity.  In cuprates and heavy fermion compounds, this issue, i.e. whether the pseudogap phase or magnetically ordered phase  terminates at a QCP inside the superconducting dome,  has been  hotly debated, but remains puzzling \cite{Coleman2005,Broun2008,Park2006,Knebel2006}.   Here we focus on  the electronic properties of the 122 family with the parent compound BaFe$_2$As$_2$, which is the most studied among the several families of iron-based superconductors discovered to date.  This family provides a so-far unique opportunity to study the evolution of the electronic properties in a wide range of the phase diagram, ranging from spin density wave (SDW) metal, through high-$T_c$ superconductor, to conventional Fermi liquid metal.  In particular, the isovalently substituted system BaFe$_2$(As$_{1-x}$P$_x$)$_2$ \cite{Kasahara2010} provides a particularly clean system because P-substitution does not induce appreciable scattering \cite{Shishido2010,vanderBeek2010}.

\section{Quantum criticality}

\subsection{Quantum phase transition}

Ordinary phase transitions are driven by thermal fluctuations and involve a change between an ordered state and a disordered state.  At absolute zero temperature, where there are no thermal fluctuations, a fundamentally new type of phase transition can occur which is called quantum phase transition \cite{Sachdev,Vojta2003,Sachdev2011,Sondhi1997,Varma2002,Lohneysen2007,Abrahams2011,Coleman2005,Si2010,Gegenwart2008}.   Quantum phase transitions are triggered by quantum fluctuations associated with Heisenberg's uncertainty principle.  This type of phase transition involves no change in entropy and can only be accessed by varying a non-thermal parameter - such as magnetic field, pressure, or chemical composition.    When the transition is continuous, the point which separates the two distinct quantum phases at zero temperature is called a QCP.  The physics of quantum criticality has become a frontier issue in condensed matter physics, in particular in strongly correlated systems.

\begin{figure}[tbh]
%\begin{center}
\includegraphics[width=0.85\linewidth]{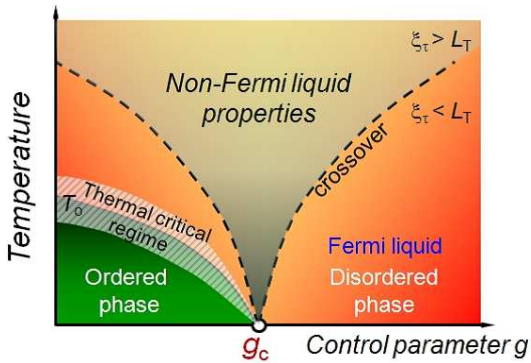}
%\includegraphics[width=0.7\linewidth]{fig1.eps}
%\end{center}
%\vspace{-8mm}
\caption{General phase diagram near a quantum critical point $g_c$. The second order phase transition  to an ordered phase at $T_o$ can be suppressed by a non-thermal parameter $g$. At finite temperatures near $T_o$, a thermal fluctuation regime exists where conventional scaling properties can be observed. When approaching the QCP from the right-hand side, the correlation length diverges. When the temperature is lowered toward zero, the thermal length diverges. Above the crossover line where these two length scales become comparable, a fan-shaped non-Fermi liquid region appears.
}
\label{fig_QCP}
\end{figure}

Figure\,\ref{fig_QCP} illustrates a typical example of the phase diagram in the vicinity of a continuous quantum phase transition.  The ground state of the system can be tuned by varying the non-thermal parameter $g$.   The system undergoes a continuous phase transition at finite temperature $T_o$.   In the close vicinity of $T_o$, there is a region of  critical thermal fluctuations.  With increasing $g$, $T_o$ decreases.  The end point of a line of this continuous finite temperature phase transition ($g=g_c$)  is the QCP, at which quantum phase transition from ordered phase to disordered phase occurs at $T=0$\,K.    At the QCP, two distinct states at $g<g_c$ and $g>g_c$ are mixed and the wavefunction is a non-trivial superposition of the two quantum states.  Approaching the QCP, the order parameter correlation length $\xi$ and correlation time $\xi_\tau$  (i.e., the correlation length along the imaginary time axis), which characterizes the dynamical (temporal) fluctuations, diverge as
\begin{equation}
\xi \propto |g-g_c|^{\nu},
\end{equation}
where $|g-g_c|$ is the distance to the QCP, and $\nu$ is the correlation length exponent, and
\begin{equation}
\xi_{\tau} \propto \xi^z,
\end{equation}
where $z$ is the dynamical exponent of the quantum phase transition \cite{Sachdev,Sondhi1997,Varma2002}.  The dispersion relation at the QCP is $\omega\propto k^z$: $z=1$ for AFM localized spin systems with spin wave excitations, and $z=2$ for itinerant AFM systems.   At finite temperature, in a quantum system, there is a finite time scale,
\begin{equation}
L_{\tau}=\frac{\hbar}{k_BT},
\end{equation}
which characterizes the thermal fluctuation (i.e., thermal length along the imaginary time axis).

%%%%%%%%%%%%%%%%%%%%%%%%%
%%%%%%%%%%%%%%%%%%%%%%%%%

The disordered phase of the system at finite temperature can be divided into distinct regimes, $L_{\tau} > \xi_{\tau}$ and $L_{\tau} < \xi_{\tau}$.  The dashed lines represent the crossover line defined by $\xi_{\tau}=L_{\tau}$ or $T\propto |g-g_c|^{\nu z}$.  For the low-temperature regime, $L_{\tau} \gg \xi_{\tau}$, the thermal time scale is much longer than the quantum time scale.  In this regime, excitations from the quantum ground state are only weakly influenced by the thermal fluctuations and hence the system can be described by the ground state wavefunction.  The quasiparticle excitations are well defined and the temperature dependence of the physical quantities can be calculated by the thermal average of independent quasiparticles.  For itinerant electron systems,  the temperature dependence of the physical quantities exhibit conventional Fermi liquid behavior, such as $T^2$-dependence of the resistivity,  $T$-independent electronic specific heat coefficient $\gamma$, where $\gamma(T) \equiv C_e(T)/T$ and $C_e(T)$ is the electronic contribution to the specific heat, and $T$-independent magnetic susceptibility $\chi$.

On the other hand, the high-temperature regime above the QCP exhibits a completely different behavior.  In this regime, where the thermal time scale is much shorter than the quantum time scale ($L_{\tau} \ll \xi_{\tau}$), the physical properties at finite temperatures are seriously influenced by the presence of the QCP at $g=g_c$ : the system at $g \neq g_c$ cannot be simply described by the ground state wave function at $g$.  In this quantum critical regime,  the temperature dependence of the physical quantities often exhibit a striking deviation from conventional Fermi liquid behavior.  For instance, in the 2D case the resistivity shows a $T$-linear behavior at low temperature, $\rho \propto T$ , $\gamma(T)$ and $\chi(T)$ become strongly temperature dependent with divergent behavior as $T \rightarrow 0$\,K, $C_e/T \propto \log T$ (for the 3D case,  $\rho \propto T^{3/2}$ and $C_e/T \propto {\rm const.}-\sqrt{T}$) \cite{Moriya2000}.

The dashed crossover lines in Figure \ref{fig_QCP} border the region of quantum critical fluctuations.  The quantum critical region has a characteristic fan shape.  Remarkably, and somewhat paradoxically, the importance of quantum criticality increases with increasing temperature, far beyond the isolated QCP at $T = 0$\,K. Thus the quantum fluctuations originated from the QCP can extend to finite temperature giving rise to unusual physical phenomena.

The quantum criticality has been studied most extensively in  non superconducting  heavy fermion systems \cite{Lohneysen2007,Si2010,Gegenwart2008}, such as CeCu$_{6-x}$Au$_x$ \cite{Schroder2000} and YbRh$_2$Si$_2$ \cite{Custers2003}, and  the ruthenate Sr$_3$Ru$_2$O$_7$ \cite{Borzi2007}.  In heavy fermion systems,  the issue of the local quantum criticality (Kondo breakdown) and quantum criticality associated with the SDW has been controversial \cite{Lohneysen2007,Si2010,Gegenwart2008}.

\subsection{QCPs in unconventional superconductors}

As mentioned above, iron-pnictides and heavy fermion compounds share common features in that unconventional  superconductivity emerges in close proximity to an antiferromagnetically ordered state and a superconducting dome appears as a function of doping or pressure, with the maximum $T_c$ found close to the extrapolated end point of the AFM transition.  The situation in the cuprates is somewhat different in that $T_c$ is small or zero close to the end point of the AFM transition. Instead, the maximal $T_c$, which occurs at a hole doping per Cu of $p=0.16$ is close to the zero temperature end point of the pseudogap phase which occurs at approximately $p=0.19$ \cite{He2011,Vishik2012,Tallon2003}.  The pseudogap is characterized by a gradual depression of the density of states at the Fermi level and a strong suppression of spin and charge excitations \cite{Timusk1999} which sets in a temperature  $T^{\ast}_{pg}$.  Recent experiments have suggested possible broken rotational and time-reversal symmetries within the pseudogap regime \cite{Lawler2010,Daou2010,Fauque2006,Xia2008} supporting the view that the pseudogap state is a distinct phase.  Within the pseudogap phase there seems to be another critical point close to $p=1/8$ where charge ordering \cite{Wu2011,Ghiringhelli2012,Chang2012} leading to Fermi surface reconstruction \cite{Vignolle2011} is observed.  There is evidence that the effective mass increases substantially close to a critical point at $p=0.10$ associated with this phase \cite{Sebastian2010}, however, the ordering that occurs here seems to correspond to a depression in $T_c$ rather than any enhancement.  So if cuprate superconductivity is also to be interpreted within the quantum critical framework it would seem that the pseudogap phase is the best candidate for the fluctuating phase.  Indeed this is also the region where the resistivity is $T$-linear over the widest temperature range.

A major open question in these three systems is whether the QCP lies beneath the superconducting dome or the criticality is avoided by the transition to the superconducting state. This question is intimately related to the following three fundamental issues.

\begin{enumerate}[ I)]
\item  Are quantum fluctuations associated with the QCP essential for superconductivity?
\item Are the non-Fermi liquid properties in the normal state above $T_c$ observed near the optimally doped regime driven by the quantum fluctuations?
\item Can unconventional superconductivity coexist with magnetic or some other exotic long range order on a microscopic level?
\end{enumerate}

This last issue is motivated by the fact that the existence of a QCP inside the dome suggests the the presence of two distinct superconducting ground states, one of which may coexist with a magnetic state.  Microscopic coexistence of superconductivity and magnetism has been a long-standing unsolved problem in heavy fermion compounds \cite{Park2006,Knebel2006,Kawasaki2003,Kawasaki2004} as well as iron pnictides \cite{Dai2012,Pratt2009,Christianson2009,Wang2011a,Iye2012,Iye2012a}.  In spite of the intensive studies using various probes, it remains unclear whether the long range AFM order truly coexists microscopically with superconducting regions or whether there is microscopic phase separation. So far the answer to this question seems to depend strongly on the experimental technique used to probe it.  The type of coexistent we are referring to here is fundamentally different to the type observed in compounds such as the  Chevrel phase, borocarbide, and some heavy fermion compounds, such as UPd$_2$Al$_3$, in which the magnetism occurs in a different electronic subsystem to the main conduction electrons \cite{Thalmeier2005}.  In the present case the same electrons are responsible for both types of behavior.

A major obstacle to probing the presence or absence of a QCP inside the superconducting dome is the presence of the superconductivity itself which makes most experimental probes insensitive to its presence.  Attempting to remove the superconductivity, for example using high magnetic field, is not straightforward either. Besides the fact that very large fields ($>50$\,T) that are required for iron-pnictides, the presence of this field will affect the original magnetic phase boundary in the zero temperature limit and may drastically change the nature of the quantum critical fluctuations.

\begin{figure}[t]
%\begin{center}
\includegraphics[width=\linewidth]{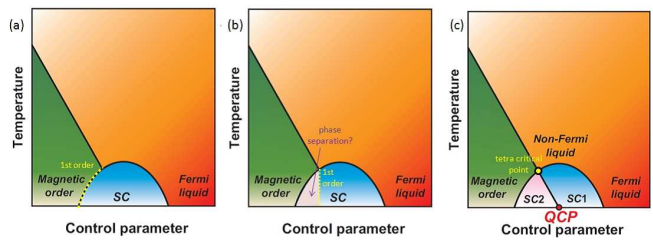}
%\end{center}
%\vspace{-8mm}
\caption{Three possible schematic phase diagrams with superconducting dome near a QCP. (a) The (magnetic) order competes with and cannot coexist with superconductivity. The boundary between the ordered phase and superconducting phase is the first order phase transition. (b) Similar to the case of (a) but the first-order nature of the boundary may lead to a sizable region of phase separation. (c) The second order phase transition line of the (magnetic) ordered phase crosses the superconducting transition line, and the QCP exists inside the superconducting dome. There should be two different phases inside the dome (SC1 and SC2). The SC2 phase is a microscopic coexistence phase of magnetic order and superconductivity.
}
\label{fig_PD}
\end{figure}

Figures\,\ref{fig_PD}(a), (b) and (c) illustrate several possible generic temperature versus non thermal control parameter phase diagrams for heavy fermion compounds and iron pnictides.
\begin{enumerate}[{Case} A:]
\item  A repulsion between AFM and superconducting (SC) order: Quantum criticality is avoided by the transition to the superconducting state (Fig.\,\ref{fig_PD}(a)).  A first-order phase transition between AFM and SC phases occurs.  There is no trace of a QCP in this case.  This phase diagram has been reported for CeIn$_3$ and  CePd$_2$Si$_2$ \cite{Kawasaki2004}.
\item  The magnetic order abruptly disappears at a temperature where magnetic and superconductivity phase boundaries meet (Fig.\,\ref{fig_PD}(b)).  A first-order or a nearly first-order phase boundary appears at a composition $x_1$ and there is no magnetic QCP. A nearly vertical first order line at $x_1$, which separates two phases, has been reported in CeRhIn$_5$ \cite{Park2006,Knebel2006}.
\item  A QCP lies beneath the superconducting dome (Fig.\,\ref{fig_PD}(c)).  The second order quantum phase transition occurs at the QCP ($x_c$) and the QCP separates two distinct superconducting phases (SC1 and SC2).  The point at which magnetic and superconductivity phase boundaries meet is a tetracritical point.  As shown later, this phase diagram is realized in BaFe$_2$(As$_{1-x}$P$_x$)$_2$ \cite{Hashimoto2012}.
\end{enumerate}

Usually when looking for a mechanism of superconductivity we think of some form of boson mediating pairing between two electrons to form a Cooper pair.  The strength and characteristic energy of the coupling then determines $T_c$. However, more generally, the transition to the superconducting state will take place when the energy of the superconducting state is lower than that of the normal state it replaces.  %In the BCS theory this occurs when the energy gain from pairing exceeds the kinetic energy cost of forming the Cooper pairs.
Therefore we can view the mechanism in which quantum criticality causes superconductivity in two different ways.  First, the quantum critical fluctuations will enhance the bosonic coupling strength and so produce strong Cooper pairing in the usual way. i.e., similar to the enhancement which occurs in electron-phonon coupled superconductors near a structural phase transition where a phonon branch softens and becomes strongly coupled to the electrons.   Second, the increase in the normal state energies caused by the quantum fluctuations will mean that a transition to the superconducting state, where such excitations are gapped out, is more energetically favorable and therefore would occur at a higher temperature than it normally would. The pairing in this case need not necessarily be solely due to the quantum fluctuations but may involve other channels such as phonons.   These two mechanisms are not be mutually exclusive but it would be natural to associate the former with case C and the latter with cases A and B. This is because when the criticality is avoided in the superconducting state (cases A and B), the superconducting gap formation surpresses the effect of quantum fluctuations on the entropy, leading to a gain in the condensation energy.

%%%%%%%%%%%%%%%%%%%%%%%%%
%%%%%%%%%%%%%%%%%%%%%%%%%

\section{122 family}

\subsection{Parent compound}

There have now been several different types of iron-pnictide superconductors discovered. The families are often abbreviated to the ratio of the elements in their parent compositions and are known as the 111, 122, 1111, 32522, 21311 types \cite{Paglione2010,Stewart2011}. In addition there are also iron-chalcogenide materials of the 11 and most recently 122 types \cite{Wen2012} which show much of the same physics as the iron-pnictides.

\begin{figure}[tbh]
%\begin{center}
\includegraphics[width=\linewidth]{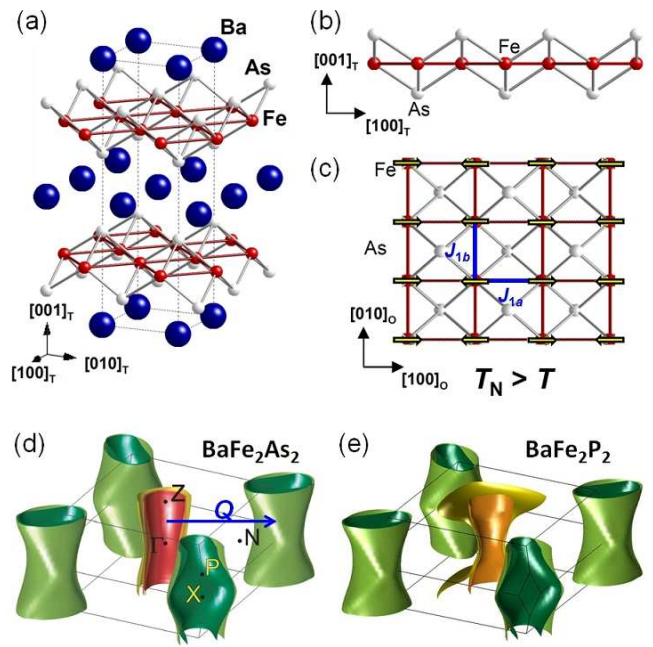}
%\includegraphics[width=0.7\linewidth]{fig3.eps}
%\end{center}
%\vspace{-8mm}
\caption{Crystal and electronic structure in BaFe$_2$As$_2$. (a) Schematic crystal structure. The dotted line represents the unit cell. The Fe-As network forms the 2D planes (b,c). The arrows in (c) illustrates the spin configuration in the antiferromagnetic state below $T_N$. (d) The Fermi surface structure of BaFe$_2$As$_2$ in the paramagnetic state. Three hole sheets near the zone center and two electron sheets near the zone corner are quasi nested when shifted by vector $\bm{Q}=(\pi,\pi,0)$. (e) For comparison the Fermi surface structure of BaFe$_2$P$_2$ is also shown. The number of hole sheets is two in BaFe$_2$P$_2$ instead of three in BaFe$_2$As$_2$, but in both cases it satisfies the compensation condition that the total volume of hole Fermi surface is the same as that of electron.
}
\label{fig_structure}
\end{figure}

Crudely the electronic and crystal structures and phase diagrams of all iron-based superconductors are quite similar.  The crystals are composed of 2D Fe layer, which is formed in a square lattice structure with an Fe-Fe distance of approximately 0.28\,nm.  The As (or P/Se/Te) atoms reside above and below the Fe layer, alternatively, and are located at the center of the Fe-atom squares, forming a tetrahedron FeAs$_4$ (Figs.\,\ref{fig_structure}(a)-(c)).  Because of the strong bonding between Fe-Fe and Fe-As sites, the geometry of the FeAs$_4$ plays a crucial role in determining the electronic properties of these systems. The Fermi surface in these materials consists of well separated hole pockets at the center of the Brillouin zone and electron pockets at the zone corners (Fig.\,\ref{fig_structure}(d)).  The parent compound is an SDW metal.   The SDW is suppressed either by chemical substitution or by pressure.  All the families exhibit a tetragonal-to-orthorhombic structural transition (i.e., broken $C_4$ symmetry) that either precedes or is coincident with the SDW transition.

%%%%%%%%%%%%%%%%%%%%%%%%%
%%%%%%%%%%%%%%%%%%%%%%%%%

\subsection{Magnetic structure}
BaFe$_2$As$_2$ undergoes  a  tetragonal-orthorhombic structural transition at $T_s=135$\,K  and at the same temperature it exhibits a paramagnetic to SDW phase transition.   The magnetic structure of BaFe$_2$As$_2$  is collinear with a small ordered moment ($\sim$ 0.9\,$\mu_B$ per Fe) \cite{Matan2009}, in which the arrangement consists of spins antiferromagnetically arranged along one chain of nearest neighbors ($a$ axis) within the iron lattice plane, and ferromagnetically arranged along the other direction ($b$ axis) (Fig.\,\ref{fig_structure}(c)).   There is a small (0.7\%) reduction in bond length along the direction where the spins are ferromagnetic coupled leading to a reduction in symmetry.  A similar collinear spin structure has also been reported  in other pnictides, such as $A$Fe$_2$As$_2$ ($A$=Ca and Sr),  $A$FeAsO (A=La, Ce, Sm, Pr, etc.) and NaFeAs, while Fe$_{1+y}$Te exhibits a bi-collinear spin structure \cite{Bao2009}.    The inelastic neutron scattering experiments have mapped out spin waves on single crystals of CaFe$_2$As$_2$, SrFe$_2$As$_2$ and BaFe$_2$As$_2$ throughout the Brillouin zone.    It has been pointed out  that neither localized nor itinerant model can satisfactorily describe these magnetic structure and excitation spectrum \cite{Dai2012}.  Recently, a possible orbital ordering has been suggested to occur together with magnetic ordering, which lifts the degeneracy between $d_{xz}$ and $d_{yz}$ orbitals \cite{Shimojima2010,Kim2011}.  Such an orbital ordered state has been suggested to be important to understand not only the magnetism \cite{Dai2012,Yildirim2009,Ku2009,Bascones2012} but also the transport properties \cite{Chen2010,Fernandes2011,Onari2012b,Phillips2012,Fernandes2012}.

\subsection{Superconductivity}

\begin{figure}[t]
%\begin{center}
\includegraphics[width=\linewidth]{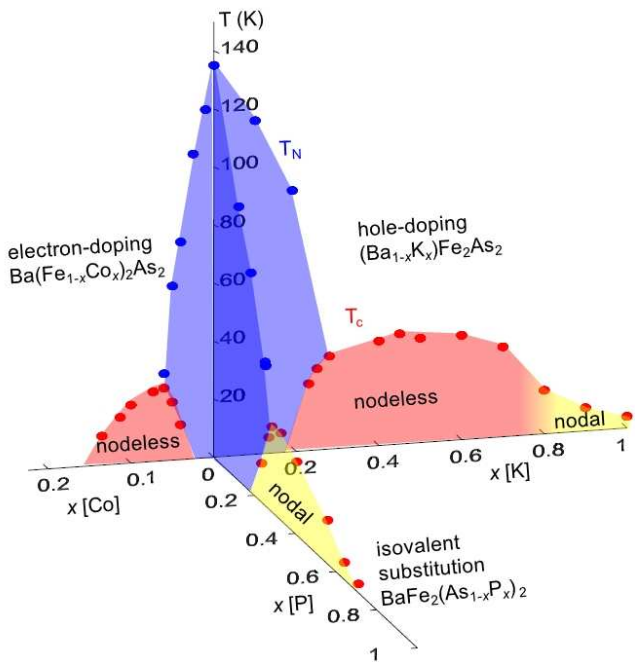}
%\includegraphics[width=0.6\linewidth]{fig4.eps}
%\end{center}
%\vspace{-8mm}
\caption{Magnetic and superconducting phase diagram of BaFe$_2$As$_2$-based materials. Superconductivity emerges when the AFM order is suppressed via either hole doping in (Ba$_{1-x}$K$_x$)Fe$_2$As$_2$ (right), electron doping in Ba(Fe$_{1-x}$Co$_x$)$_2$As$_2$ (left), or isovalent substitution in BaFe$_2$(As$_{1-x}$P$_x$)$_2$ (bottom). In the P substituted system and in the overdoped region of K-doped system, the superconducting gap has line nodes.
}
\label{fig_122}
\end{figure}

High-temperature superconductivity develops when the `parent' AFM/orthorhombic phase is suppressed, typically by introduction of dopant atoms.  In most of the iron-based compounds, the magnetic and structural transition temperatures split with doping \cite{Paglione2010,Stewart2011}.  The hole doping is achieved by substitution of Ba$^{2+}$ by K$^{1+}$ in (Ba$_{1-x}$K$_x$)Fe$_2$As$_2$ and electron doping by substitution of Fe by Co  in Ba(Fe$_{1-x}$Co$_x$)$_2$As$_2$ or Ni in Ba(Fe$_{1-x}$Ni$_x$)$_2$As$_2$.   High-$T_c$ superconductivity appears even for the isovalent doping with phosphorous in BaFe$_2$(As$_{1-x}$P$_x$)$_2$ or ruthenium in Ba(Fe$_{1-x}$Ru$_x$)$_2$As$_2$.  The magnetic and superconducting phase diagram of the BaFe$_2$As$_2$-based systems is shown in Fig.\,\ref{fig_122}. In the hole doped (Ba$_{1-x}$K$_x$)Fe$_2$As$_2$, the structural/magnetic phase transition crosses the superconducting dome at $x\sim 0.3$ and a maximum $T_c$ of 38\,K appears at $x\simeq 0.45$.  Upon hole doping, the hole pocket expands and the electron pocket shrinks and disappears at $x\sim 0.6$ \cite{Malaeb2012}.   The superconductivity is observed even at the hole-doped end material ($x = 1$), KFe$_2$As$_2$, which corresponds to 0.5 holes/Fe atom.  In the electron doped Ba(Fe$_{1-x}$Co$_x$)$_2$As$_2$, the maximum $T_c$ of 22\,K appears at $x=0.07$.  In contrast to the hole-doped case, superconductivity vanishes at only 0.15 electrons/Fe atom (although note that each doped K atom only adds 0.5 holes per Fe). This electron-hole asymmetry in the phase diagram has been attributed to an enhanced Fermi surface nesting in the hole-doped compounds.  The superconductivity of $T_c=31$\,K appears in heavily-electron doped $A_x$Fe$_{2-y}$Se$_2$ with no hole pockets \cite{Wen2012}.

%%%%%%%%%%%%%%%%%%%%%%%%%
%%%%%%%%%%%%%%%%%%%%%%%%%

\subsection{Isovalent substitution system}
The isovalently `doped' BaFe$_2$(As$_{1-x}$P$_x$)$_2$ system is a particularly suitable to study the detailed evolution of the electronic properties because of the following reasons.   In isovalent `doping' with no introduction of additional charge carriers,  the dopant changes the electronic structure mainly because of differences in ion size.  In fact, the phase diagram of BaFe$_2$(As$_{1-x}$P$_x$)$_2$  can be retraced with hydrostatic pressure from any starting P concentration \cite{Kasahara2010}, suggesting that pressure is somehow equivalent to substitution.  Observation of quantum oscillations  in a wide $x$-range  ($0.38 \leq x \leq  1$) demonstrates the low scattering rate of the defects introduced by P substitution \cite{Shishido2010,Analytis2010,Arnold2011,Walmsley}, particularly for the electron sheets.

According to density function theory (DFT) band-structure calculations \cite{Kasahara2010,Shishido2010}, three hole sheets exist around the zone center ($\Gamma$ point) in BaFe$_2$As$_2$, while one of them is absent in BaFe$_2$P$_2$ (Figs.\,\ref{fig_structure}(d) and (e)). Both compounds have two electron pockets around the zone corner ($X$ point).  The three dimensionality of the hole Fermi surfaces is quite sensitive to the pnictogen position $z_{Pn}$. The substitution of P for As reduces both the $c$ axis length and $z_{Pn}$ and eventually leads to the loss of one of the hole sheets and a strong increase in the warping of another which gains strong $d_{z^2}$ character close to the top of the zone (Z point). This increased Fermi surface warping upon doping weaken the nesting along the $(\pi,\pi)$ direction.  In contrast to the significant changes in the hole sheets, the electron sheets are almost unchanged in the calculations, although experimentally a significant reduction in their volume is found \cite{Shishido2010}.

\subsection{Superconducting gap structure and symmetry}
Detailed knowledge of the superconducting gap structure and how it varies between different families can be useful in helping to decide between microscopic theories \cite{Hirschfeld2011,Mazin2008,Kuroki2009,Graser2009,Ikeda2010,Chubukov2012,Kontani2010,Onari2012,Thomale2011}.   The superconducting gap structure in the 122 family has been studied extensively by means of various experimental techniques \cite{Hirschfeld2011}.  Fully gapped superconductivity has been well established in the optimally doped regime of electron-doped Ba(Fe$_{1-x}$Co$_x$)$_2$As$_2$ \cite{Tanatar2010} and hole-doped (Ba$_{1-x}$K$_x$)Fe$_2$As$_2$ \cite{Ding2008,Hashimoto2010a}, indicating $A_{1g}$ ($s$-wave) symmetry.

%Fully gapped superconductivity is also reported in heavily electron doped $A_x$Fe$_{2-y}$Se$_2$. In this system, the fact that there is no node in the small electron pocket around the $\Gamma$-point excludes  $d$-wave symmetry \cite{Xu2012}, indicating a gap function with $A_{1g}$ symmetry.

On the other hand, the presence of line nodes have been reported in heavily hole doped KFe$_2$As$_2$ \cite{Fukazawa2009,Dong2010,Hashimoto2010b,Reid2012,Okazaki2012} and throughout the whole superconducting region of the phase diagram in isovalently doped BaFe$_2$(As$_{1-x}$P$_x$)$_2$  \cite{Hashimoto2010,Yamashita2011,Hashimoto2012,Wang2011}.
%For KFe$_2$As$_2$ with three hole pockets around $\Gamma$-point,  recent ARPES measurements \cite{Okazaki2012} reported that while eight line nodes appears in the middle hole band, the inner hole pocket is essentially fully gapped, indicating the gap function with $A_{1g}$ symmetry.
  For BaFe$_2$(As$_{1-x}$P$_x$)$_2$, there is no evidence of vertical line nodes in the hole pockets located at the zone center.  Although the position of the line nodes in this system is controversial \cite{Shimojima2011,Yoshida2013,Yamashita2011,Zhang2012}, it is very likely that the gap function has $A_{1g}$ symmetry.  Recent results detailing the effect of electron irradiation on the magnetic penetration depth demonstrate that the line nodes are lifted by the impurities \cite{Mizukami2013} indicating that they are not symmetry protected \cite{Mishra2009}.  From the above we conclude that the gap structure is not universal, but the gap symmetry is universal, i.e. $A_{1g}$ symmetry, at least in the 122 family.

\section{QCP hidden beneath the Superconducting Dome}

The method of isovalent substitution offers an ideal route to quantum criticality \cite{Abrahams2011,Dai2009}, as distinct from charge carrier doping or application of external pressure.   Since BaFe$_2$As$_2$ exhibits SDW order and BaFe$_2$P$_2$ does not, we can place the two end materials on either sides of $g_c$ along the tuning parameter axis in the phase diagram. As $x$ increases, the tuning parameter $g$ increases.

%%%%%%%%%%%%%%%%%%%%%%%%%
\begin{figure}[t]
%\begin{center}
\includegraphics[width=\linewidth]{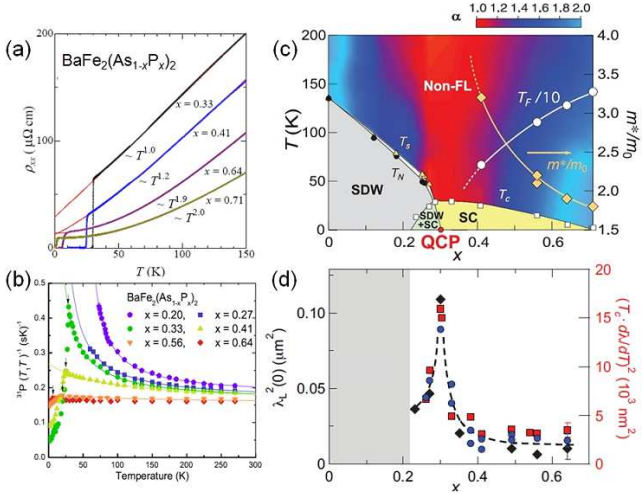}
%\includegraphics[width=0.9\linewidth]{fig5.eps}
%\end{center}
%\vspace{-8mm}
\caption{Evidence for the QCP in the superconducting dome in BaFe$_2$(As$_{1-x}$P$_x$)$_2$. (a) Temperature dependence of in-plane resistivity $\rho_{xx}$ for $0.33\le 0.71$ \cite{Kasahara2010}. The red lines are the fit of normal-state $\rho_{xx}(T)$ to power-law dependence $\rho_0+AT^\alpha$ (Eq.\,(4)). (b) Temperature dependence of NMR $1/T_1T$ measured for the $^{31}$P nuclei for several compositions \cite{Nakai2010}. The lines are the fit to the Curie-Weiss temperature dependence Eq.\,(5). (c) Phase diagram composed from a color plot of the exponent $\alpha$ of the temperature dependence of the resistivity. The effective mass $m^*$ and Fermi temperature $T_F$ extracted from the quantum oscillation measurements are also plotted. (d) The $x$ dependence of the square of zero-temperature London penetration depth $\lambda_L^2(0)$  \cite{Hashimoto2012} determined by the Al-coated method (diamonds), surface impedance (circles), and slope of the temperature dependence of $\delta\lambda_L(T)$ (squares, right axis).
}
\label{fig_AsP}
\end{figure}
%%%%%%%%%%%%%%%%%%%%%%%%%

\subsection {Transport properties}
As shown in Fig.\,\ref{fig_AsP}(a), at $x=0.30$ where the maximum $T_c$ is achieved, the in-plane resistivity $\rho$ shows linear temperature dependence, $\rho=\rho_0+AT$ \cite{Kasahara2010}, which is a hallmark of non-Fermi liquid behavior.   We note that along with  the $T$-linear resistivity, a striking enhancement of  Hall coefficient at low temperatures and apparent violation of Kohler's law in magnetoresistance have been reported \cite{Kasahara2010},  which are also indicative of non-Fermi liquid behavior \cite{Nakajima2007}.   On the other hand, at $x\gtrsim 0.6$,  the resistivity follows the Fermi liquid relation of $\rho=\rho_0+AT^2$ (Fig.\,\ref{fig_AsP}(a)).   The color shading in Fig.\,\ref{fig_AsP}(c) represents the value of the resistivity exponent in the relation
\begin{equation}
\rho=\rho_0+AT^{\alpha}.
\end{equation}
A crossover from non-Fermi liquid to Fermi liquid with doping is clearly seen.  The region of the phase diagram,  which includes a funnel of $T$-linear resistivity centered on $x\approx 0.3$ shown by red,  bears a striking resemblance to the quantum critical regime shown in Fig.\,\ref{fig_QCP}, although the superconducting dome masks the low temperature region.   Thus the transport results suggest  the presence of QCP at $x\approx0.3$.

\subsection {Magnetic properties}
The nuclear magnetic resonance (NMR) experiments give important information about the low-energy magnetic excitations of the system.  The Knight shift $K$ and spin-lattice relaxation rate $1/T_1$ of BaFe$_2$(As$_{1-x}$P$_x$)$_2$  have been measured with various P concentrations \cite{Nakai2010}.  $K$ is almost $T$-independent for all $x$, indicating that the density of states (DOS) does not change substantially with temperature.  The $^{31}$P relaxation rate $1/T_1$ is sensitive to the AFM fluctuations: $1/T_1T$ is proportional to the average of the imaginary part of the dynamical susceptibility $\chi(\bm{q},\omega_0)/\omega_0$, $1/T_1T \propto \Sigma_q|A(\bm{q})|^2 \chi'' (\bm{q},\omega_0)/\omega_0$, where $A(\bm{q})$ is the hyperfine coupling between $^{31}$P nuclear spin and the surrounding electrons and $\omega_0$ is the NMR frequency.  In the Fermi liquid state, the Korringa relation $T_1TK^2={\rm const.}$ holds, but it fails in the presence of strong magnetic fluctuations.  In particular, AFM correlations enhance $1/T_1$ through the enhancement of $\chi(\bm{q}\neq 0$), without appreciable change of $K$.

At $x=0.64$,  $1/T_1T$ is nearly temperature independent (Fig.\,\ref{fig_AsP}(b)), indicating the Korringa relation $T_1TK^2={\rm const.}$  This Fermi liquid behavior in the magnetic properties is consistent with the transport properties.  As $x$ is varied towards the optimally doping, $1/T_1T$ shows a strong temperature dependence, indicating a dramatic enhancement of the AFM fluctuations.    It has been reported that in the paramagnetic regime $T_1$ is well fitted with the 2D AFM spin fluctuation theory of a nearly AFM metal,
\begin{equation}
\frac{1}{T_1T}=a+\frac{b}{T+\theta},
\end{equation}
where $a$ and $b$ are fitting parameters and $\theta$ is the Curie-Weiss temperature  (Fig.\,\ref{fig_AsP}(b)).   Nakai \textit{et al.} \cite{Nakai2010} reported that while $a$ and $b$ change little with $x$,  $\theta$  exhibits a strong $x$ dependence.  With decreasing $x$, $\theta$ decreases and goes to zero at the critical concentration $x\approx0.3$, where non-Fermi liquid behavior in the resistivity is observed.  At a second order AFM critical point, the singular part of $1/T_1T$ is expected to vary $1/T$, i.e. $\theta=0$  because the dynamical susceptibility diverges at $T=0$\,K, or the magnetic correlation length continue to increases down to 0\,K.   Therefore the NMR results also suggest a QCP at $x\approx0.3$.

\subsection{Fermi surface and mass renormalization}

Discovering how the Fermi surface evolves as the material is tuned from a non-superconducting conventional metal (right-hand side of the phase diagram) toward the non-Fermi liquid regime near the SDW phase boundary is an important step toward gaining a complete understanding of the mechanism that drives high-$T_c$ superconductivity.   Quantum oscillations arise from the Landau quantization of the energy levels of metals in high magnetic fields and can be used to map out the detailed Fermi-surface structure. They are usually observed at very low temperatures and in very clean single crystals.   The frequencies of the observed oscillations, $F$, (as a function of inverse magnetic field) provide very accurate measurements of the Fermi surface extremal cross-sectional areas, $A_k$, via the Onsager relation, $F=(\hbar/2\pi e)A_k$.  Moreover, the quasiparticle effective masses ($m^{\ast}$) on each of the extremal orbits,  which are important for understanding  the degree of electronic correlation,  can be extracted from the temperature-dependent amplitude of the oscillations.  The quantum oscillations observed in the magnetization or torque are known as the de Haas van Alphen effect (dHvA) effect.  In BaFe$_2$(As$_{1-x}$P$_x$)$_2$ dHvA oscillations have been observed in a wide doping range \cite{Shishido2010}, indicating that the substitution of As by P does not induce appreciable scattering.

For the end member BaFe$_2$P$_2$, the dHvA oscillations originating from all the Fermi surface sheets are observed and hence the complete Fermi surface is precisely determined \cite{Arnold2011}.  All the orbits have relatively uniform mass enhancements $m^{\ast}/m_b$ ranging from 1.6 to 1.9 ($m_b$ is the DFT band mass).   For the As-substituted samples, the dHvA oscillations from the hole sheets are rapidly attenuated with decreasing $x$ \cite{Shishido2010}: signals from the hole sheets have been reported only up to $x=0.63$ \cite{Analytis2010}, still some way from the SDW phase boundary, at which $T_c$ reaches its maximum.   On the other hand, the dHvA quantum oscillations from the electron sheets, in particular the signals from the $\beta$-orbits on the outer electron sheet, have been observed in a wide doping range up to $x=0.38$ ($T_c=28$\,K), which is fairly close to the boundary \cite{Shishido2010,Walmsley}.

The quasiparticle effective masses exhibit a steep upturn in samples of progressively lower P-concentration $x$ (Fig.\ \ref{fig_AsP}).  The enhanced mass of the electron sheet at $x=0.38$ reported by the ARPES \cite{Yoshida2011} is quantitatively consistent with dHvA results.    Concomitantly with the mass enhancement,  the volume of the electron sheets (and via charge neutrality also the hole sheets) shrinks linearly, which is not expected from the DFT calculations.   The inferred  Fermi temperature $T_F=\hbar eF/m^{\ast}k_B$ decreases rapidly with decreasing $x$ (Fig.\,\ref{fig_AsP}(c)).    It is highly unlikely that these changes are a simple consequence of the one-electron band structure but instead they likely originate from many-body interactions.   It is generally believed that strong quantum fluctuations near the QCP lead to a notable many-body effect, which seriously modify the quasiparticle masses.  Therefore the enhancement in $m^{\ast}$ and shrinkage of Fermi surface, and hence precipitous drop of Fermi temperature as the material is tuned toward the magnetic order phase boundary,  suggest a QCP is being approached.  Thus the electronic structure revealed by dHvA experiments is consistent with a QCP near the SDW boundary.

\subsection{Thermodynamic evidence of the QCP}

The enhancement of the quasiparticle mass leads to the enhancement of $\gamma$ determined by the thermodynamic specific heat.  Unfortunately, owing to the high superconducting transition  temperatures,   the direct determination of $\gamma$ is extremely difficult.  However, $\gamma$ can be calculated by the jump of the specific heat at $T_c$ by the relation,
\begin{equation}
\gamma=\Delta C/\alpha_c T_c,
\end{equation}
 by making the reasonable assumption that $\alpha_c$, which takes the value $1.43$ for weak coupling $s$-wave superconductors, does not change appreciably near the QCP.   Very recent systematic study of $\Delta C$ measurements using very high-quality single crystals \cite{Walmsley} revealed a striking enhancement of the quasiparticle mass as the magnetic order phase boundary is approached, which is \textit{quantitatively} consistent with the mass determined by both the dHvA measurements and the superfluid density.

\subsection{Superfluid density}

Despite the bulk thermodynamic and transport signatures of quantum critical behavior of the normal quasiparticles at finite temperature, they are not sufficient to pin down the location the QCP because of the overlying superconducting dome.   Due to the above mentioned potential problem associated with destroying the superconductivity with a strong field, a direct probe which can trace across the QCP at zero temperature in zero field is desirable.

One property which directly probes the superconducting state well below $T_c$ at zero field is the magnetic penetration depth $\lambda_L$.  In the clean, local (London) limit, the absolute value of $\lambda_L$ in the zero temperature limit is given by the following expression
\begin{equation}
    \lambda_{L_x}^{-2}(0)=\frac{\mu_0e^2}{4\pi^3\hbar}\oint dS \frac{v_x^2}{|v|}
\end{equation}
where $\oint dS$ refers to an integral over the whole Fermi surface, and $v_x$ and $\lambda_{L_x}$ are the $x$ components of the Fermi velocity $v$ and the penetration depth respectively. In a multiband system with simple Fermi surface, this may be simplified to
\begin{equation}
\lambda_L^{-2}(0)=\mu_0e^2\sum_i n_i /m^{\ast}_i,
\end{equation}
where $n_i$ and $m^{\ast}_i$ are the number density (proportional to the volume) and average mass of the carriers in band $i$ respectively.  The key point here is that $\lambda_L$ is a direct probe of the normal state properties of the electrons which form the superconducting state. Measurements on very high-quality crystals are indispensable because impurities and inhomogeneity may otherwise wipe out the signatures of the quantum phase transition.

Figure\,\ref{fig_AsP}(d) shows the doping dependence of the squared in-plane London penetration length $\lambda_L^2(0)$ in the zero-temperature limit, which are determined by three different methods \cite{Hashimoto2012}. The first is the lower-$T_c$ superconducting film coating method using  a high precision tunnel diode oscillator  (operating frequency of $\sim 13$\,MHz) \cite{Prozorov2000,Gordon2010}.    The second is the microwave cavity perturbation technique by using a superconducting resonator (resonant frequency 28\,GHz) and a rutile cavity resonator  (5\,GHz), both of which have very high quality factor $Q\sim10^6$.  In the third method, $\lambda_L(0)$ is determined by the slope of $T$-linear dependence of  $\delta\lambda_L(T)=\lambda_L(T)-\lambda_L(0)$, which is determined by the tunnel diode oscillator, by assuming that the gap structure evolves weakly across the phase diagram and the $x$ dependence of $d\lambda_L/d(T/T_c)$ will mirror that of  $\lambda_L(0)$ .

All three methods give very similar $x$ dependencies.  The most notable feature is the sharp peak in  $\lambda_L^2(0)$ at $x=0.30$.  This striking enhancement on approaching $x=0.30$ from \emph{either} side is naturally attributed to the critical fluctuations associated with a second-order quantum phase transition, providing strong evidence for the presence of a QCP at $x=0.30$.     It should be noted that in a Galilean invariant system such an enhancement of $\lambda_L(0)$ is not expected because the self-energy renormalization effect is canceled out by the so called backflow correction \cite{Leggett1965}.  However, recent theoretical work \cite{Levchenko2012} has argued that  in multiband systems (such as the iron-pnictides),  Galilean invariance is broken and then electron correlation effects give rise to a striking enhancement of the superfluid electron mass and hence the magnetic penetration depth has a peak due to the mass enhancement at the QCP.  In addition, the presence of nodes leads to a further enhancement of $\lambda_L(0)$ at the QCP.   Moreover, it has been suggested that  a strong renormalization of effective Fermi velocity due to quantum fluctuations occurs only for momenta $\bm{k}$ close to the nodes in the superconducting energy gap $\Delta(\bm{k})$ \cite{Hashimoto2013}.  This ``nodal quantum criticality'' is expected to lead to a peculiar $T^{3/2}$-dependence of $\lambda_L$ near the QCP.  Such a $T$-dependence has been reported not only in BaFe$_2$(As$_{1-x}$P$_x$)$_2$ with $x$ close to 0.3 but also in $\kappa$-(BEDT-TTF)$_2$Cu(NCS)$_2$ and CeCoIn$_5$, which may be also close to a QCP.

\subsection{Continuous  quantum phase transition inside the dome}

The penetration depth measurements provide clear and direct evidence of the presence of QCP lying beneath the superconducting dome \cite{Hashimoto2012}.  This implies that the non-Fermi liquid behavior indicated by the red region in Fig.\,\ref{fig_122}(a) is most likely associated with the finite temperature quantum critical region linked to the QCP.  In addition, the enhanced quasiparticle mass implies that the Fermi energy is suppressed, which is usually less advantageous for high $T_c$.  The fact that the highest $T_c$ is nevertheless attained right at the QCP with the most enhanced mass strongly suggests that the quantum critical fluctuations help to enhance superconductivity in this system.

Moreover, this transition immediately indicates two distinct superconducting ground states.  The strong temperature dependence of $\delta\lambda_L(T)$ at low temperatures observed on both sides of the QCP argues against a drastic change in the superconducting gap structure \cite{Hirschfeld2011,Fernandes2010}. The fact that the zero-temperature extrapolation of the AFM transition $T_N(x)$ into the dome coincides with the location of the QCP leads us to conclude that the QCP separates a pure superconducting phase and a superconducting phase coexisting with the SDW order (Fig.\,\ref{fig_PD}(c)). The present results strongly suggest that superconductivity and SDW coexist on a microscopic level, but compete for the same electrons in the underdoped region. This competition is evidenced by the overall larger $\lambda_L(0)$ values in the SDW side of the QCP than the other side (Fig.\,\ref{fig_122}(d)), which might be explained by a smaller Fermi surface volume due to partial SDW gapping.  The microscopic coexistence is also supported by the enhancement of $\lambda_L^2(0)$ on approaching the QCP from the SDW side, which is not expected in the case of phase separation.

We stress that the observed critical behavior of $\lambda_L(0)$ has never been reported in any other superconductors, including other iron-based \cite{Gordon2010,Luan2011}, heavy-fermion and cuprate \cite{Tallon2003} superconductors.  The doping evolution of $\lambda_L(0)$ has been reported in electron doped Ba(Fe$_{1-x}$Co$_x$)$_2$As$_2$, but $\lambda_L(0)$ increases monotonically with decreasing $x$ and no special feature is observed even when acrossing the magnetic phase boundary at $x\approx 0.06$.  There are several possible reasons for this.  Firstly, recent neutron diffraction measurements on electron-doped Ba(Fe$_{1-x}$Ni$_x$)$_2$As$_2$ reported that the commensurate static AFM order changes into transversely incommensurate short-range AFM order near optimal superconductivity \cite{Luo2012}, implying that the first order magnetic transition takes place in the phase diagram and that there is no QCP in the electron doped compound, although there are some reports suggesting the quantum critical behavior in the normal-state properties of Ba(Fe$_{1-x}$Co$_x$)$_2$As$_2$ \cite{Ning2010,Yoshizawa2012,Chu2012}.  Secondly, the QCP anomaly may be smeared out by a greater degree of the electronic disorder caused by Co doping in the Fe planes.  Thus  it remains unclear whether the long-range AFM order truly coexists microscopically with superconducting regions in electron doped system.

The microscopic coexistence of SDW and superconductivity is also supported by $^{31}$P-NMR measurements in BaFe$_2$(As$_{1-x}$P$_x$)$_2$ with $x=0.25$ \cite{Iye2012}.  The magnetic moment of Fe atom grows rapidly as the temperature is decreased below $T_N$, but it is seriously reduced when the system undergoes the superconducting transition at $T_c$, indicating a direct coupling between magnetic and superconducting order parameters. These results appear to indicate that the electrons on the same Fermi surfaces contribute to both magnetic ordering and unconventional superconductivity in  BaFe$_2$(As$_{1-x}$P$_x$)$_2$.   It should be noted that in heavy fermion compounds CeCo(In$_{1-x}$Cd$_x$)$_5$ and CeRhIn$_5$, where $T_N$ is higher than $T_c$, such a reduction is not observed.  These results  suggest that magnetism and superconductivity may emerge from different parts of the Fermi sheets in these heavy fermion compounds, implying weak coupling between magnetism and superconductivity, which is consistent with the phase separation shown in Fig.\,\ref{fig_PD}(b).  In cuprates, a QCP associated with the pseudogap formation has been discussed at the putative critical hole concentration $p_0 \sim 0.19$ inside the superconducting dome.  It should be noted, however,  that  a broad minimum of $\lambda_L^2(0)$ reported in Bi$_2$Sr$_2$CaCu$_2$O$_{8+x}$ at $p_0$ \cite{Tallon2003} is in sharp contrast to the striking enhancement of $\lambda_L^2(0)$  in BaFe$_2$(As$_{1-x}$P$_x$)$_2$.  Therefore, the nature of the QCP in cuprates, if present, may be very different from that in iron-pnictides.

\subsection{Nematic quantum criticality}
In iron-pnictides orbital physics and magnetism are highly entangled due to the strong interaction between spin and orbital motion, similar to manganites \cite{Kugel}.  Closely related with this issue is the electronic nematicity, which is a unidirectional self-organized state that breaks the rotational symmetry of the underlying lattice.  The nematicity and its relation to superconductivity have been one of the important issues in iron pnictides \cite{Yildirim2009,Ku2009,Chen2010,Chuang2010,Chu2010,Shimojima2010,Yi2011,Kim2011,Fernandes2011,Fernandes2012,Fernandes2012a,Kontani2011a,Onari2012b,Bascones2012,Phillips2012,Kasahara2012,Blomberg2013}.

It has been suggested that the electronic nematicity in pnictides is associated with the tetragonal-to-orthorhombic structural transition.    As the magnetic critical point at $x_c=0.30$ is approached the structural transition temperature $T_s$ decreases along with the SDW ordering temperature.  Therefore the nematic fluctuations are expected to be enhanced near $x_c$ \cite{Fernandes2012,Kontani2011a,Onari2012}.  Indeed, the critical behavior of the nematic fluctuations have been reported in ultrasound \cite{Yoshizawa2012} and elastic response of resistivity anisotropy measurements \cite{Chu2012} for Ba(Fe$_{1-x}$Co$_x$)$_2$As$_2$.

Recent magnetic torque experiments in BaFe$_2$(As$_{1-x}$P$_x$)$_2$ suggest that the electronic nematicity appears at a temperature $T^*$ much higher than $T_s$ and that the superconducting dome is covered under the $T^*$ line \cite{Kasahara2012}.  Moreover, very recent ARPES measurements show that the inequivalent energy shifts of  $d_{xz}$ and $d_{yz}$ bands, which have been observed in the AFM state below $T_N$ \cite{Shimojima2010,Yi2011}, seem to emerge at around this $T^*$ line \cite{Shimojima2013}.  In addition,  the NMR $1/T_1T$  becomes enhanced below $T^*$\cite{Nakai2013}.   Although further studies are necessary to clarify the nature of the electronic state changes at $T^*$ and $T_s$,  an interesting possibility is that a ferro orbital ordering occurs at $T_s$,  while an antiferro orbital ordering takes place at $T^*$ \cite{Kontani2011a}.  In this case, the associated anomalies, which are clearly detected at $T_s$,  may not be detected at $T^*$  by long wave-length ($\bm{q}\sim0$) probes such as elastic constants.

\section{BCS-BEC crossover}

%%%%%%%%%%%%%%%%%%%%%%%%%
\begin{figure}[t]
%\begin{center}
\includegraphics[width=0.8\linewidth]{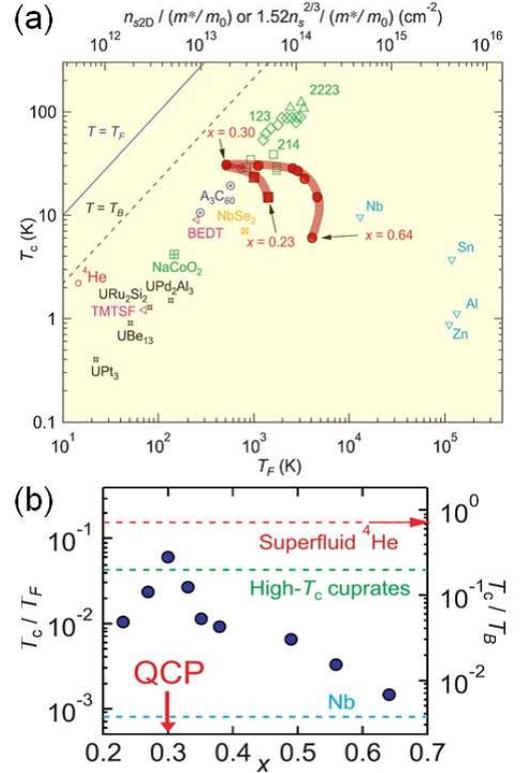}
%\includegraphics[width=0.55\linewidth]{fig6.eps}
%\end{center}
%\vspace{-8mm}
\caption{Comparisons between the superconducting transition temperature $T_c$ and the effective Fermi temperature $T_F$ \cite{Hashimoto2012}. (a) The so-called Uemura plot, where $T_c$ is plotted against $T_F$ estimated from the superfluid density. The data for BaFe$_2$(As$_{1-x}$P$_x$)$_2$ with different $x$ (red circles) show a quite different behavior from the linear relation found for high-$T_c$ cuprates and other exotic superconductors, and rather bridge the gap between conventional BCS superconductors and unconventional superconductors. The red think lines are guides for the eyes. (b) $T_c/T_F$ ratio as a function of $x$. The dashed lines represent typical values of $T_c/T_F$ for Nb (blue), cuprates (green) and of $T_c/T_B$ for superfluid $^4He$ (red).
}
\label{fig_Uemura}
\end{figure}
%%%%%%%%%%%%%%%%%%%%%%%%%

The discovery of the QCP inside the dome brings another new aspect of the superconducting state in iron-pnictides, which has never been realized in any other superconductors.  The red symbols in Fig.\,\ref{fig_Uemura}(a) show $T_c$ plotted as a function of $T_F$ in BaFe$_2$(As$_{1-x}$P$_x$)$_2$ for various $x$. Because the relevant Fermi surface sheets are nearly cylindrical, $T_F$ may be estimated directly from the superfluid density $\lambda_L^{-2}(0)$ via the relation, $T_F=(\hbar^2 \pi)n_{s\mathrm{2D}}/k_Bm^{\ast}\approx \hbar^2 \pi/ \mu_0e^2d\lambda_L^2(0)$, where $n_{s\mathrm{2D}}$ is the carrier concentration within the superconducting planes and $d$ is the interlayer spacing.  The results of various superconductors are also shown in Fig.\,\ref{fig_Uemura}(a), where $T_F$ is given by $T_F=(\hbar^2/2)(3\pi^2)^{2/3}n_s^{2/3}/k_Bm^{\ast}$ for 3D systems \cite{Uemura2004}. The dashed line in Fig.\,\ref{fig_Uemura}(a) corresponds to the Bose-Einstein condensation (BEC) temperature for ideal 3D boson gas, $T_B=\frac{\hbar^2}{2 \pi m^{\ast}k_B}(\frac{n_s}{2.612})^{2/3}=0.0176T_F$. In a quasi-2D system, this value of $T_B$ provides an estimate of the maximum condensate temperature. The evolution of the superfluid density in the present system is in sharp contrast to that in cuprates, in which $T_c$ is roughly scaled by $T_F$. Figure\,\ref{fig_Uemura}(b) depicts the P-composition dependence of $T_c$ normalized by the Fermi (or BEC) temperature, $T_c/T_F$ ($T_c/T_B$). In the large composition region ($x>0.6$), $T_c/T_F$ is very small, comparable to that of the conventional superconductor Nb.  As $x$ is decreased, $T_c/T_F$ increases rapidly, and then decreases in the SDW region after reaching the maximum at the QCP ($x=0.30$). What is remarkable is that the magnitude of $T_c/T_B (\approx 0.30)$ at the QCP exceeds that of cuprates and reaches as large as nearly 40\% of the value of superfluid $^4$He. This sharp peak in $T_c/T_F$ implies that the pairing interaction becomes strongest at the QCP and that the quantum critical fluctuations help to enhance superconductivity in this system.

\section{Conclusions}

In this review we have discussed the quantum criticality of  iron-based high-$T_c$  superconductors, addressing the issue of a QCP lying beneath the superconducting dome,  which we believe to be crucially important for understanding of anomalous  non-Fermi liquid  properties, microscopic coexistence between superconductivity and magnetic order, and the mechanism of superconductivity.  In cuprates and heavy fermion compounds, this issue, i.e. whether the pseudogap phase in cuprates and SDW phase in heavy fermion compounds terminate at a QCP deep inside the dome,   has been hotly debate, but remains puzzling.

We have shown that the isovalent doped pnicitide BaFe$_2$(As$_{1-x}$P$_x$)$_2$ is an ideal system to study this issue, because we can tune the electronic properties in a wide range of the phase diagram, ranging from SDW metal, through high-$T_c$ superconductor, to conventional Fermi liquid metal without introducing appreciable scattering.  The transport properties, NMR, quantum oscillatons, and specific heat, all suggest the presence of the QCP at $x_c=0.30$.  Moreover, zero-temperature London penetration depth measurements provide clear and direct evidence of the QCP lying beneath the superconducting dome.  The QCP inside the dome includes the following important implications.
\begin{itemize}
\item  The QCP is the origin of the non-Fermi liquid behavior above $T_c$.
\item Unconventional superconductivity coexists with a spin density wave antiferromagnetism on a microscopic level.
\item  The quantum critical fluctuations help to enhance the high-$T_c$ superconductivity.
\end{itemize}

The presence or absence of a QCP inside the dome is still a question of debate not only in cuprates and heavy fermions but also in other pnictides.  With three unconventional superconducting systems to compare and contrast,  the vital clues that could be used to solve the mystery of anomalous electronic properties and unconventional superconductivity might be uncovered.

\section*{Acknowledgments}
The authors acknowledge collaborations with A.\ F. Bangura, A.\,E. B\"ohmer, K. cho, A.\,I. Coldea, H. Eisaki, A. Fujimori, H. Fukazawa, R.\,W. Giannetta, K. Hashimoto, H. Ikeda, S. Kasahara, H. Kitano, K. Ishida, T. Iye, A. Iyo, L. Malone, C. Meingast, Y. Mizukami, Y. Nakai, K. Okazaki, C. Proust, R. Prozorov, C. Putzke, N. Salovich, T. Shimojima, S. Shin, H. Shishido, M.\,A. Tanatar, T. Terashima, S. Tonegawa, M. Yamashita, D. Vignolles, P. Walmsley, D. Watanabe, and T. Yoshida. 
We also thank the following for helpful discussions: E. Abrahams, R. Arita, A.\,V. Chubukov, I. Eremin, D.\,L. Feng, R.\,M. Fernandes, T. Hanaguri, P.\,J. Hirschfeld, H. Kontani, K. Kuroki, I\,.I Mazin, S. Sachdev, J. Schmalian, Q. Si, T. Tohyama, Y.\,J. Uemura, S. Uchida, and H.\,H. Wen.

\end{document}